Comment on
"The effect of variable viscosity on the flow and heat transfer on a continuous stretching surface" by A. Hassanien [ ZAMM, 1997, Vol. 77, pp. 876-880]

Asterios Pantokratoras
Associate Professor of Fluid Mechanics
School of Engineering, Democritus University of Thrace,
67100 Xanthi – Greece
e-mail:apantokr@civil.duth.gr

## 1. INTRODUCTION

The problem of forced convection along an isothermal, constantly moving plate is a classical problem of fluid mechanics that has been solved for the first time in 1961 by Sakiadis (1961). Thereafter, many solutions have been obtained for different aspects of this class of boundary layer problems. Solutions have been appeared including mass transfer, varying plate velocity, varying plate temperature, fluid injection and fluid suction at the plate. The work by Hassanien (1997) belongs to the above class of problems, including a linearly varying velocity and the variation of fluid viscosity with temperature. The author obtained similarity solutions considering that viscosity varies as an inverse function of temperature. However, the Prandtl number, which is a function of viscosity, has been considered constant across the boundary layer. It has been already confirmed in the literature that the assumption of constant Prandtl number leads to unrealistic results (Pantokratoras, 2004, 2005). The objective of the present paper is to obtain results considering both viscosity and Prandtl number variable across the boundary layer. As will be shown later the differences of the two methods are very large in some cases.

## 2. THE MATHEMATICAL MODEL

Consider the flow along a moving plate placed in a calm environment with u and v denoting respectively the velocity components in the x and y direction, where x is the coordinate along the plate and y is the coordinate perpendicular to x. For steady, two-dimensional flow the boundary layer equations including variable viscosity are

continuity equation: $$\frac{\partial u}{\partial x} + \frac{\partial v}{\partial y} = 0 \qquad (1)$$

momentum equation: $$u\frac{\partial u}{\partial x} + v\frac{\partial v}{\partial y} = \frac{1}{\rho_a}\frac{\partial}{\partial y}\left(\mu\frac{\partial u}{\partial y}\right) \qquad (2)$$

energy equation:
$$u\frac{\partial T}{\partial x} + v\frac{\partial T}{\partial y} = \alpha\frac{\partial^2 T}{\partial y^2} \qquad (3)$$

where T is the fluid temperature, $\mu$ is the dynamic viscosity, $\alpha$ is the thermal diffusivity, and $\rho_a$ is the ambient fluid density.

The boundary conditions are as follows:

at $y = 0$    $U=bx$, v=0, $T=T_w$ (4)

as $y \rightarrow \infty$    u =0, $T = T_a$ (5)

where $T_w$ is the plate temperature, $T_a$ is the ambient fluid temperature and bx is the velocity of the moving surface.

The viscosity is assumed to be an inverse linear function of temperature given by the following equation (Hassanien 1997)

$$\frac{1}{\mu} = \frac{1}{\mu_a}[1 + \gamma(T - T_a)] \qquad (6)$$

where $\mu_a$ is the ambient fluid dynamic viscosity and $\gamma$ is a thermal property of the fluid. Equation (6) can be rewritten a follows

$$\frac{1}{\mu} = c(T - T_e) \qquad (7)$$

where c=$\gamma/\mu_a$ and $T_e = T_a - 1/\gamma$ are constants and their values depend on the reference state and the thermal property of the fluid.

The equations (1), (2) and (3) form a parabolic system and were solved directly, without any transformation, by a method described by Patankar (1980). The finite difference method is used with primitive variables x, y and a space marching procedure is used in x direction with an expanding grid. A detailed description of the solution procedure may be found in Pantokratoras (2002) where all fluid properties ( viscosity, thermal diffusivity and density) have been considered as functions of temperature.

## 3. RESULTS AND DISCUSSION

The most important quantities for this problem are the wall heat transfer and the wall shear stress defined as

$$\theta'(0) = \frac{x}{T_w - T_a}\mathrm{Re}^{-1/2}\left[\frac{\partial T}{\partial y}\right]_{y=0} \qquad (8)$$

$$f''(0) = \frac{\vartheta_e - 1}{\vartheta_e} \frac{\mu_w}{\rho_a U^2} Re^{1/2} \left[ \frac{\partial u}{\partial y} \right]_{y=0} \tag{9}$$

where $\theta$ is the dimensionless temperature $(T - T_a)/(T_w - T_a)$ and f is the dimensionless stream function for which the following equation is valid

$$f' = \frac{u}{U} \tag{10}$$

The Reynolds number is defined as

$$Re = \frac{Ux}{\nu_a} \tag{11}$$

and $\theta_e$ is a constant defined by

$$\theta_e = \frac{T_e - T_a}{T_w - T_a} = -\frac{1}{\gamma(T_w - T_a)} \tag{12}$$

In equations (8), (9) and (10) the prime represents differentiation with respect to similarity variable $\eta$ defined as

$$\eta = \frac{y}{x} Re^{1/2} = y \left( \frac{b}{\nu_a} \right)^{1/2} \tag{13}$$

It should be mentioned here that when $\theta_e \rightarrow \infty$ the fluid viscosity becomes equal to ambient viscosity.

In order to test the accuracy of the present method, results were compared with those available in the literature. The wall heat transfer $\theta'(0)$ and the wall shear stress $f''(0)$ for the present problem with constant viscosity and Pr $=0.7$ are $-0.454449$ and $-1.0$ respectively (Abo-Eldahab and El Aziz 2004). The corresponding quantities calculated by the present method are $-0.4541$ and $-1.0$. The comparison is satisfactory and this happens for other Pr numbers.

In contrast to the above direct solution of equations (1), (2) and (3), Hassanien (1997) transformed these equations into the following similarity equations

$$f''' - \frac{1}{\vartheta - \vartheta_e} f'' \vartheta' - \frac{\vartheta - \vartheta_e}{\vartheta_e} (ff'' - f'^2) = 0 \tag{14}$$

$$\vartheta'' + Pr f\vartheta' = 0 \tag{15}$$

It should be mentioned here that in the transformed energy equation (15) the Prandtl number has been assumed constant across the boundary layer. Hassanien (1997) have calculated this Prandtl number at ambient temperature from the following equation

$$\Pr_a = \frac{\nu_a}{\alpha} \tag{16}$$

However, the Prandtl number is a function of viscosity and as viscosity varies across the boundary layer, the Prandtl number varies, too. As will be shown below this assumption leads to unrealistic results in some cases.

In table 1 the wall shear stress and the wall heat transfer are given for ambient Prandtl number 0.7. In the table the results by Hassanien (1997) have been also included for comparison. In the last column of the table the Prandtl number at the plate ($\Pr_w$) is included.

Table 1. Values of f'(0) and θ'(0) for $\Pr_a$=0.7

| $\theta_e$ | f'(0) | | | θ'(0) | | | |
|---|---|---|---|---|---|---|---|
| | Present Work | Hassanien (1997) | Difference % | Present Work | Hassanien (1997) | Difference % | $\Pr_w$ |
| -10 | -1.0644 | -1.0561 | <1 | -0.4466 | -0.4487 | <1 | 0.64 |
| -8 | -1.0775 | -1.0696 | <1 | -0.4447 | -0.4447 | <1 | 0.62 |
| -6 | -1.0992 | -1.0916 | <1 | -0.4417 | -0.4442 | <1 | 0.60 |
| -4 | -1.1414 | -1.1401 | <1 | -0.4360 | -0.4408 | 1 | 0.56 |
| -2 | -1.2579 | -1.2593 | <1 | -0.4198 | -0.4417 | 5 | 0.47 |
| -1 | -1.4592 | -1.4536 | <1 | -0.3922 | -0.3980 | 1 | 0.35 |
| -0.25 | -2.2892 | -2.2856 | <1 | -0.2907 | -0.3140 | 8 | 0.14 |
| -0.1 | -3.3655 | -3.3603 | <1 | -0.2078 | -0.2491 | 20 | 0.06 |
| -0.05 | -4.6250 | -4.6204 | <1 | -0.1601 | -0.2096 | 31 | 0.03 |
| -0.01 | -10.0875 | -10.0794 | <1 | -0.0825 | -0.1601 | 94 | 0.007 |
| 2 | -0.6502 | -0.6466 | <1 | -0.4996 | -0.5247 | 5 | 1.40 |
| 4 | -0.8467 | -0.8442 | <1 | -0.4752 | -0.5059 | 6 | 0.93 |
| 6 | -0.9047 | -0.8979 | <1 | -0.4676 | -0.4691 | <1 | 0.84 |
| 8 | -0.9320 | -0.9246 | <1 | -0.4641 | -0.4657 | <1 | 0.80 |
| 10 | -0.9460 | -0.9402 | <1 | -0.4620 | -0.4637 | <1 | 0.78 |

From table 1 it is seen that the wall shear stress, calculated by the two methods, are in agreement. For wall heat transfer things are different. For large values of the parameter $|\theta_e|$ the results are in agreement but as $|\theta_e|$ decreases the results of the two methods diverge. It is advocated here that the wall heat transfer for $Pr_a=0.7$ and low $|\theta_e|$ calculated by Hassanien are unrealistic. In figure 1 the temperature profiles are shown for $Pr_a=0.7$ and $\theta_e =-0.01$ and $-0.025$. It is seen that the real temperature profiles, calculated with variable Pr number, are much wider than those given by Hassanien (1997) calculated with constant ambient Pr number. The error is introduced by considering that the ambient Pr number is valid in the entire boundary layer but this is valid only for large values of $|\theta_e|$. For low values of of $|\theta_e|$ the real Pr number inside the boundary layer is much smaller than the ambient one as it is shown in figure 2. From table 1 we see that for $\theta_e =-0.01$ the Prandtl number at the plate is 100 times smaller than the ambient one. It is well known in the boundary layer theory that large Pr numbers correspond to "narrow" temperature profiles and small Pr numbers to wider temperature profiles. This is the reason for the difference in the temperature profiles and the wall heat transfer between the two methods.

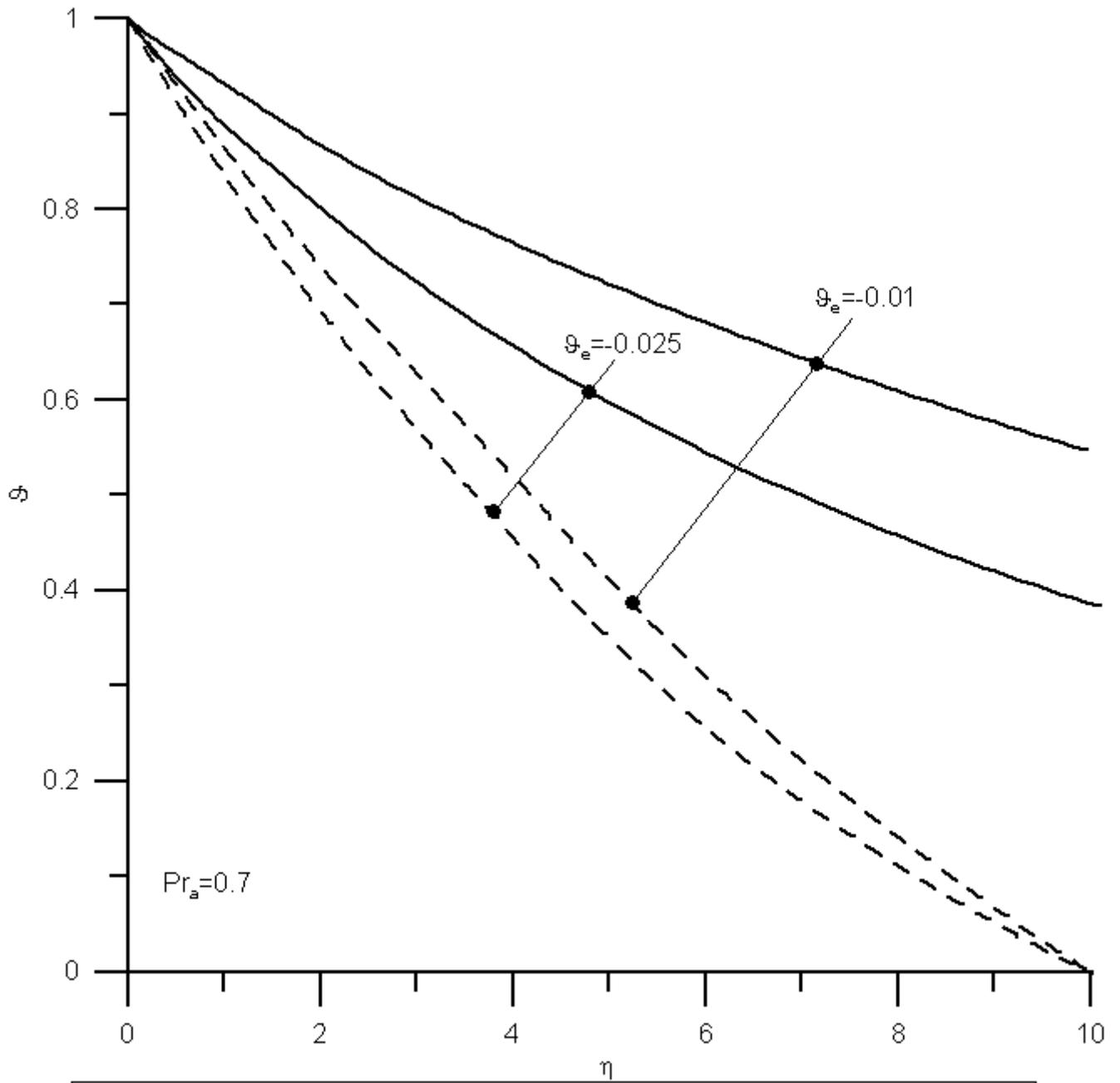

Figure 1. Temperature distribution for ambient Prandtl number 0.7 and $\vartheta_e$ = -0.01 and -0.025:solid line, present work : dashed line, Hassanien (1997).

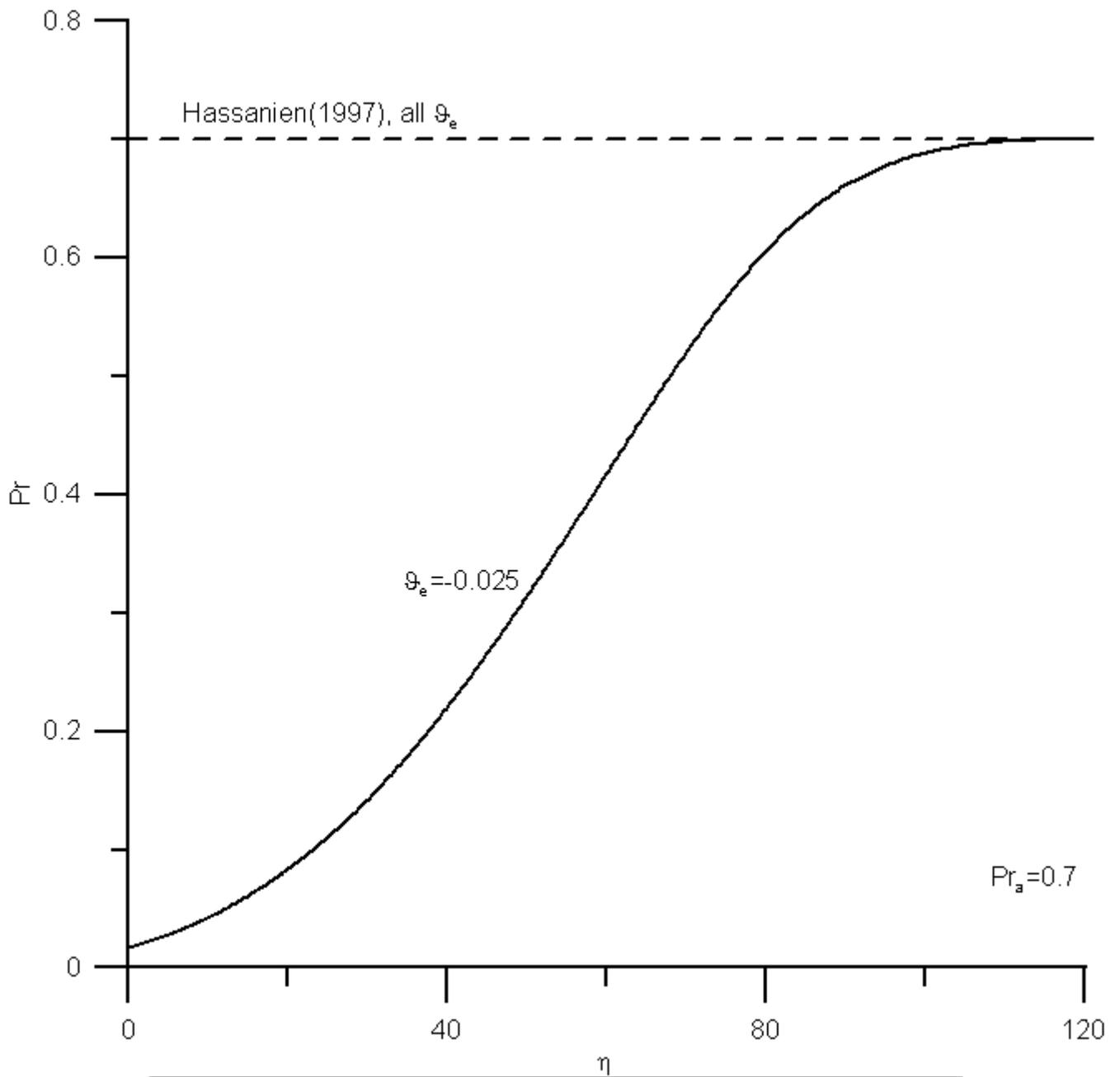

Figure 2. Distribution of Prandtl number across the boundary layer for $Pr_a = 0.7$: solid line, present work; dashed line, Hassanien (1997).